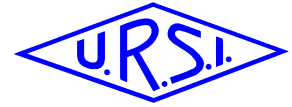

# Properties of DH Type II Radio Bursts and Their Space Weather Implications

N. Gopalswamy[1] and P. Mäkelä[1,2]
(1)NASA Goddard Space Flight Center, Greenbelt, Maryland, USA
(2)The Catholic University of America, Washington DC, USA
nat.gopalswamy@nasa.gov

## Abstract

We report on the properties of type II radio bursts observed by the Radio and Plasma Wave Experiment (WAVES) on board the Wind spacecraft over the past two solar cycles. We confirm that the associated coronal mass ejections (CMEs) are fast and wide, more than half the CMEs being halos. About half of the type II bursts extend down to 0.5 MHz, corresponding to a heliocentric distance of tens of solar radii. The DH type II bursts are mostly confined to the active region belt and their occurrence rate follows the solar activity cycle. Type II burst occurring on the western hemisphere of the Sun and extending to lower frequencies are good indicators of a solar energetic particle event.

## 1. Introduction

The Radio and Plasma Wave Experiment (WAVES [1]) on board the Wind spacecraft has been providing information on low-frequency radio emission from the Sun since its launch in 1994. For example, WAVES helps track shocks emitting type II radio bursts from the Sun all the way to Earth. Most of the previous radio instruments observed at frequencies below 2 MHz [2], leaving a large gap between the lower frequency limit of ground based observations (~20 MHz) and space-based observations (<2 MHz). The gap corresponds to the heliocentric range of 2-10 solar radii (Rs) where many interesting phenomena take place. The Solar and Heliospheric Observatory (SOHO) was launched in 1995 and observes the corona up to a heliocentric distance of ~32 Rs [3]. Currently both SOHO and Wind are located at Sun-Earth L1 point and observe overlapping sections of the corona. While the SOHO field of view ends at ~32 Rs, WAVES can continue to track shocks all the way to 1 AU. Observing the same region of the corona by the two instruments resulted in a number of discoveries such as interacting CMEs [4], shock formation at large distances from the corona [5,6], rapid change in the power law index describing the density fall off with distance [7], ending frequencies of type IV radio bursts [7], hierarchical relationship between CME kinetic energy and the wavelength range of type II bursts [8], the universal bandwidth characteristic [9], the close connection to solar energetic particle (SEP) events [10-11], and the relation to sustained gamma-ray emission (SGRE) [12]. One of the aspects highly relevant to space weather is the fact that the decameter-hectometric (DH) type II bursts can identify fast and wide CMEs that propagate far into the interplanetary medium [13-14].

In this paper, we update the properties of DH type II bursts, the associated CMEs and SEP events, taking advantage of the extended data available over two solar cycles.

## 2. Data Analysis and Results

We consider all DH type II bursts observed by the Wind/WAVES experiment from the beginning of solar cycle 23 (May 1996). Any type II burst starting at frequencies 1-14 MHz are included as DH type II bursts. These include purely DH type II bursts and the bursts extending to kilometric wavelengths. The bursts are already listed at the CDAW data Center (https://cdaw.gsfc.nasa.gov/CME_list/radio/waves_type2.html) along with the associated CMEs, flares, and the source location of the eruption. The starting and ending frequencies are also listed in this catalog. For the present work, we use the sky-plane speed and width in obtaining statistical properties.

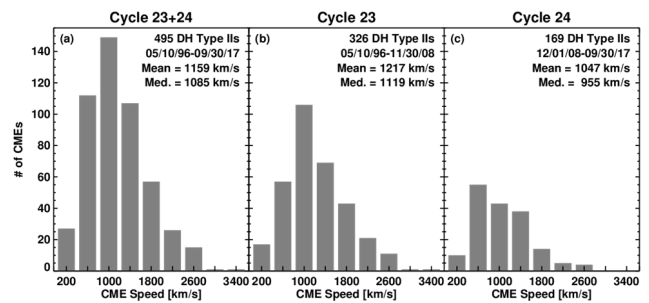

**Figure 1**. Speed distributions of CMEs associated with DH type II bursts in (a) cycles 23 and 24 combined, (b) cycle 23 and (c) cycle 24. The speeds are measured in the sky plane, so the true speeds are likely to be higher.

2.1 CME Kinematics
In an earlier study of ~100 DH type II bursts that occurred over the first five years of observations, it was found that these bursts are caused by energetic CMEs: the average CME speed was 961 km/s and the width of non-halo CMEs (width <200 deg.) was 102 deg. in the sky plane [13]. For limb events, the numbers were 1144 km/s and 112 deg.

Fifty two of the 101 CMEs (or 52%) were halos. Over the past 24 years, the number of DH type II bursts has increased five-fold. Figure 1 shows CME speed distributions for cycle 23 and 24 and the combined set. The average speeds are similar in the three plots and in agreement with the earlier study [13]. Figure 2 shows the width distributions, again for cycles 23 and 24 and the combined set. Overall, the halo CME fraction has slightly increased to 57.5%. The halo fraction in cycle 24 (62.1%) is higher than that in cycle 23 (54.6%) probably due to the higher halo CME abundance in cycle 24 because of the reduced total pressure in the heliosphere [15-16]. The halo fraction over the full cycle 23 did not change much from the initial study [13].

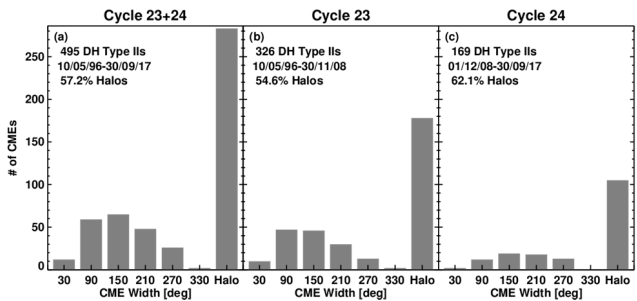

**Figure 2.** Width distributions of CMEs associated with DH type II bursts in (a) cycles 23 and 24, (b) cycle 23 and (c) cycle 24. The widths are measured in the sky plane. The fraction of halos in each population is shown on the plots.

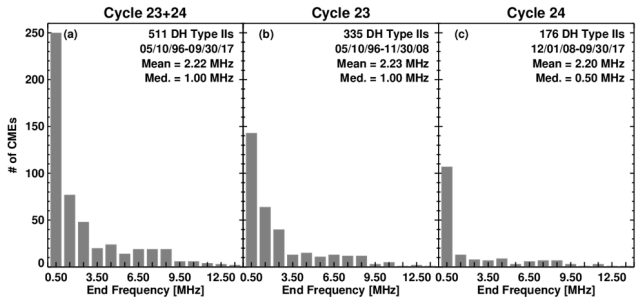

**Figure 3.** Distributions of the ending frequencies of DH type II bursts in (a) cycles 23 and 24 combined, (b) cycle 23 and (c) cycle 24. Lower frequency emission originates from larger distances from the Sun.

2.2 Type II Ending Frequencies
Figure 3 shows the distributions of ending frequencies of DH type II bursts. About half of the 511 DH type II bursts (49%) observed in cycles 23 and 24 combined had ending frequency below 0.5 MHz. This fraction was only 42% in all of cycle 23 and 60% in all of cycle 24 (until the end of September 2017). This means the cycle-24 shocks survived over a larger distance from the Sun than those in cycle 23. This may also reflect the reduced Alfven speed in the ambient medium inferred from solar wind data [15-16].

2.3 Solar Cycle Variation of the Occurrence Rate
Figure 4 shows the number of DH type II bursts in each Carrington rotation since 1996 until the end of September 2017 in comparison with the sunspot number (SSN) averaged over Carrington rotation period. There is clearly an overall correlation between SSN and the occurrence rate of DH type II bursts per Carrington rotation. Even the double peak of the cycle-24 SSN is also present in the DH type II burst rate. Solar cycle 24 started in December 2008 and is still in progress. The current length of cycle 24 is 106 months long. For comparing the number of events between cycles 23 and 24, we also count the number of DH type II bursts over the first 106 months in cycle 23. The number of DH type II bursts were 286 and 176 in cycles 23 and 24, respectively over the first 106 months. It is clear that the DH type II activity declined in cycle 24 by 38%. Over the same period, the SSN declined by 44%. On the other hand, the number of fast and wide CMEs declined from 427 in cycle 23 to 248 in cycle 24, which is a 42% decrease. Noting that there was a 4-month data gap in cycle 23, the corrected decline is ~44%. All these numbers are within 6%, suggesting the decline in solar activity as the main reason for the reduced DH type II activity.

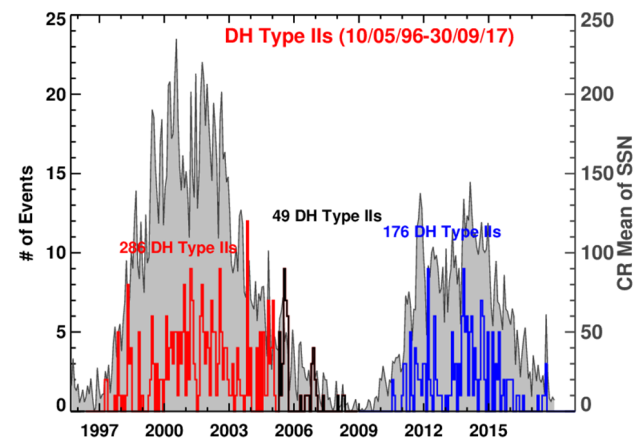

**Figure 4.** Occurrence rate of DH type II radio bursts summed over Carrington rotation periods plotted as a function of time. Cycle 23 and 24 events are in red and blue colors, respectively. Cycle 24 is ongoing with 176 bursts from 2008 December 1 to 2017 September 30. Over the same epoch, cycle 23 has 286 bursts. During the remainder of cycle 23, there are 49 bursts. The sunspot number (SSN) averaged over Carrington rotation periods is shown in gray.

2.4 DH Type II Radio Bursts and SEP Association
One of the main motivation for studying type II radio bursts is their close physical connection to SEP events: the same shock accelerates electrons that produce type II radio bursts and protons (and heavier ions) that are detected as SEP events when the particles have magnetic connectivity to the observer. It was shown earlier that there is a good correlation between the number of SEP events and the western hemispheric type II bursts [17]. Here we consider the solar source distributions of the DH type II bursts in Fig. 4 and consider their SEP association. The number of SEP events (152) in cycles 23 and 24 is much smaller than that of the DH type II bursts over the same period (511). In order to understand this, we have plotted the solar sources of 495 DH type II bursts whose sources are on the disk (see

Fig. 5). Notice that the sources are predominantly located in the latitude range of ±30 deg., corresponding to the active region belt. This is because type II bursts in the DH domain require fast and wide CMEs that can only be powered by the high magnetic energy stores in active region magnetic fields. This is consistent with the overall correlation between the SSN and the occurrence rate of DH type II bursts shown in Fig. 4.

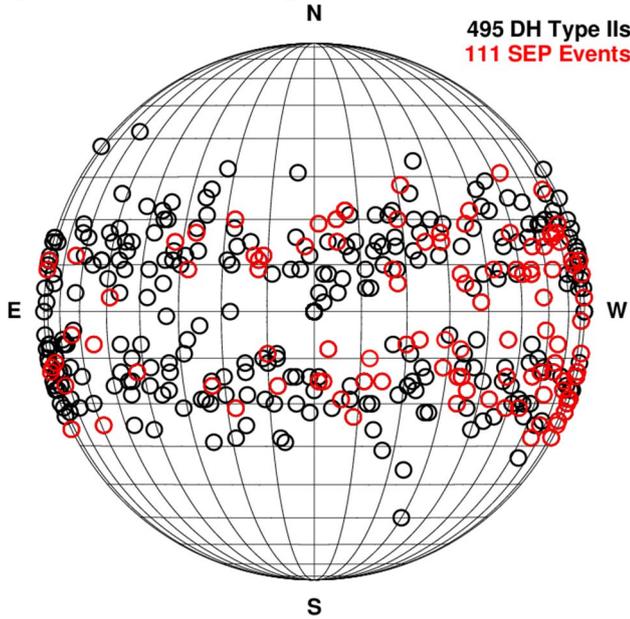

**Figure 5.** Source locations of CMEs responsible for type II radio bursts observed by Wind/WAVES. The data points in red correspond to those type II bursts that are associated with a large SEP event as detected by one of the GOES satellites. The grid lines on the solar disk are at every 10 deg.

In Fig. 5, the type II bursts associated with large SEP events are distinguished by red symbols. Only 111 of the 495 DH type II bursts (or 22%) were associated with large SEP events. Over the same period, there were 152 large SEP events. There are several reasons for this.

1. The DH type II bursts include all events that have type II emission at frequencies below 14 MHz. Many type II bursts appear only briefly. The association rate of SEP events increases when the wavelength range of type II bursts increases [8, 11]. This is because the CME kinetic energy organizes the wavelength range over which type II emission occurs [18]. In fact, there is one-to-one correspondence between large SEP events and type II bursts with emission components at all plasma frequencies between the Sun and Earth. This point can be illustrated using the ending frequency distributions in Fig. 3. Consider the 250 events in the 0.5 MHz bin. Since the distribution of the DH type II burst sources are uniform across the disk, we expect roughly 125 sources on the western hemisphere. Excluding the 26 eastern-hemispheric SEP events from the total of 152, we have 126 SEP events from the western hemisphere, roughly matching the number of DH type II bursts from the western hemisphere. We do not expect perfect match because type II bursts not reaching 0.5 MHz have SEP association.

2. Many type II events occur in quick succession and SEP events last for days. This results in many SEP events occurring against a high background from previous events and hence are not distinguished as new events.

3. We have counted only large SEP events in which the >10 MeV proton flux exceeds 10 pfu. There are many SEP events with >10 MeV flux occurring between 0.1 pfu (typical GOES background level) and 10 pfu.

4. The shock needs to be well connected to the observer to detect an SEP event. On the other hand, the type II bursts can be observed from eruptions anywhere on the disk and even from behind the limb to some extent.

Table 1. Large SEP events from the east limb

| CME Date | CME UT | Source Location | $V_{CME}$ km/s | Flare size | SEP Intensity |
|---|---|---|---|---|---|
| 2001/12/28 | 20:30 | S26E90 | 2216 | X3.4 | 76 pfu |
| 2002/07/20 | 22:06 | S13E90 | 1941 | X3.3 | 28 pfu |
| 2005/07/27 | 04:54 | N11E90 | 1787 | M3.7 | 41 pfu |
| 2011/09/22 | 10:48 | N09E89 | 1095 | X1.4 | 35 pfu |
| 2013/06/21 | 03:12 | S16E73 | 1900 | M2.9 | 14 pfu |
| 2014/02/25 | 01:25 | S12E82 | 2147 | X4.9 | 24 pfu |

The importance of connectivity is clear from the fact that most of the SEP-associated type II bursts originate from western hemisphere. Out of the 111 SEP events plotted in Fig. 5, only 26 are from the eastern hemisphere (23%). Only seven of the 111 SEP events (or ~6%) are from close to the limb (east of E70). SEP events from the east limb are due to exceptionally fast CMEs. One of the seven events occurred on 2005 September 7, but there was no CME information because of a coronagraph data gap. The event was very intense (the >10 MeV peak flux was ~1880 pfu). The associated flare was extremely large (X17), which indicates an extreme CME. The remaining six east-limb SEP events associated with DH type II bursts and CMEs are listed in Table 1. Clearly, the CME speeds are very high, averaging to ~2000 km/s. Because of the high speeds, the shock flanks are still fast enough to accelerate particles in sufficient numbers to be detected as a large SEP event at Earth when the western flank crosses the Sun-Earth field lines. The SEP peak intensity was small in these events ranging from 14 to 76 pfu with an average of 36 pfu (the 2005 September 7 event is an exception). It must be noted that the SEP intensity is expected to be much higher at observing locations well connected to these eruptions. In fact, the cycle-24 events from the eastern hemisphere were well-connected to STEREO-B, which detected large SEP events. For example, the 2011 September 22 SEP event had a >10 MeV flux exceeding 1000 pfu at STEREO-B [19]. Similarly, the 2014 February 25 event had a >10 MeV peak flux of ~400 pfu. This event is also a hard-spectrum event because Fermi satellite's Large Area Telescope (Fermi/LAT) detected an SGRE event in the energy range >100 MeV. The gamma-rays require >300 MeV protons that propagate sunward from the shock to produce SGRE via the pion decay mechanism [12, 20].

## 3. Summary and Conclusions

We investigated the properties of more than 500 DH type II bursts and those of the associated CMEs. The CMEs are fast (>1000 km/s) and wide (>50% are halos) on average. In about half of the events, ending frequency was at or below 0.5 MHz, suggesting strong shocks. The occurrence rate of DH type II bursts roughly follows the sunspot number. Accordingly, the solar sources of these bursts are mostly located in the active region belt. Sources of type II bursts associated with large SEP events are generally found on the western hemisphere because such sources are well connected to an observer along the Sun-Earth line. In summary, observing a type II burst at frequencies below 14 MHz is a sure way of telling that a CME-driven shock is accelerating electrons. It is generally the case that the same shock accelerates electrons and ions. If the type II burst occurs on the western hemisphere, there is a good chance that an SEP event is observed. The chance rapidly increases when the CME speed increases or equivalently, the radio emission extends over wide range of wavelengths. We find that the ending frequency of DH type II bursts is a good indicator if the presence of a large SEP event. However, the plasma frequency of ~0.5 MHz occurs at tens of solar radii from the Sun, so it is not very useful in predicting an SEP event. On the other hand the close correspondence between type II bursts and SEP events provides a better understanding of CME-driven shocks that accelerate electrons and protons. One useful information is the increased chance of western hemispheric CMEs producing a large SEP event when the initial speed of the CME is high (>1000 km/s). It may be possible to develop a scheme to quickly estimate the CME speed to determine whether the resulting SEP event will be large. Another possibility is the see if the DH type II burst continues to lower frequencies, say about 5 MHz, it may be an indication of a large SEP event.

## 5. Acknowledgements

We benefited from the open data policy of *SOHO*, *STEREO*, *GOES*, and *Wind* teams. Work supported by NASA's LWS and Heliophysics GI programs.